\documentclass[11pt,reqno]{amsart} 
\calclayout

\usepackage{amsthm}
\usepackage{amsmath}
\usepackage{amsfonts}
\usepackage{amssymb}
\usepackage[dvips,all]{xypic}
\usepackage[dvips,all]{xy}
\usepackage{graphicx}
\usepackage{url}
\usepackage{enumerate}
\usepackage{lscape}
\usepackage[usenames,dvipsnames]{color}
\usepackage{multicol}
\usepackage{mathtools}
\usepackage{array}
\usepackage{blkarray}
\usepackage{subfig}

\usepackage{tikz}

\usepackage{xcolor}

\theoremstyle{definition} 
\newtheorem{theorem}{Theorem}[section]

\newtheorem{example}[theorem]{Example}

\newtheorem*{question*}{Question} % not numbered
\newtheorem*{problem*}{Problem} % not numbered
\theoremstyle{remark}

\numberwithin{equation}{section}

\newcommand{\Ftot}{F_{\mbox{tot}}}
\newcommand{\Stot}{S_{\mbox{tot}}}
\newcommand{\Ktot}{K_{\mbox{tot}}}
\newcommand{\Ptot}{F_{\mbox{tot}}}

\newcommand{\R}{\mathbb{R}}

\newcommand{\lra}{\leftrightarrows}

\definecolor{dgreen}{rgb}{.2,.6,.2}
\colorlet{darkgreen}{black!30!dgreen}

\definecolor{dblue}{rgb}{0.0,0.0,0.68}

\usepackage{lineno}
\begin{document} 

\title[Post-translational modification systems]{Dynamics of post-translational modification systems: \\ recent progress and future directions}
\author{Carsten Conradi and Anne Shiu}
{\address{	
	CC: Dept.\ of Life Science Engineering, HTW Berlin, Germany;
    AS: Dept.\ of Mathematics, Mailstop 3368, Texas A\&M Univ., College Station TX~77843--3368, USA
}}
\email{carsten.conradi@htw-berlin.de,annejls@math.tamu.edu}

% for arXiv
\date{24 October 2017}
%\date{\today}

%---------------------------------
%: ABSTRACT
%---------------------------------
\begin{abstract}
Post-translational modification (PTM) of proteins is important for signal transduction, and hence significant effort has gone toward understanding how PTM networks process information.  
This involves, on the theory side, analyzing the dynamical systems arising from such networks. Which networks are, for instance, bistable?  Which networks admit sustained oscillations?  Which parameter values enable such behaviors?  
In this Perspective, we highlight recent progress in this area and point out some important future directions. 
Along the way, we summarize several techniques for analyzing general networks,
such as eliminating variables to obtain steady-state parametrizations, 
and harnessing results on how incorporating intermediates affects dynamics.
% on dynamics and steady states.
%How do cells process information?   How do signaling networks work?  Here we consider these questions in the context of common modules within signaling networks, specifically, post-translational modification systems.  Indeed, it is well known that post-translational modification, such as phosphorylation, acetylation, and methylation, form key components of signaling networks, including MAPK cascades.
\end{abstract}

\maketitle

%---------------------------------
%: INTRODUCTION
%---------------------------------
\section{Introduction}
Post-translational modification (PTM) of proteins (for instance, by
phosphorylation or methylation) %acetylation  
plays a key role 
in signal transduction, with disruptions having
implications for human health~\cite{cohen-role,Anas}. 
Much research 
therefore has focused on determining how networks of PTM's
process and encode information~\cite{escape,SK}. 
The goal, on the theory side, is to understand the dynamical systems
arising from such networks.  Indeed, qualitative properties,
such as the capacity for bistability or oscillations, 
  may indicate that the underlying biochemical mechanism acts as a %can realize a
  biological switch or clock \cite{tyson-albert}. 
Bistability~\cite{bistab-cascade,mapk-bistable} and oscillations~\cite{yeast-mapk-oscillations,oscillations-mapk-cancer}, have been observed in cellular signaling networks, such as 
mitogen-activated protein kinase (MAPK) cascades, in which PTM networks play a key role.
% Phosphorylation plays a key role in cellular signaling networks, such as 
%{\em mitogen-activated protein kinase (MAPK) cascades}, which enable cells to make decisions (to differentiate, proliferate, die, and so on)~\cite{chang-karin}.  
%This decision-making role of MAPK cascades % in decision-making %and committing to 
%decisions 
%suggests that they exhibit switch-like behavior, i.e., bistability.  Indeed, this has been seen in experiments .  Sustained oscillations also have been observed~\cite{yeast-mapk-oscillations,oscillations-mapk-cancer}, hinting at a role in keeping time. % or synchronization.

Although PTM networks are often large, with 
parameter values that are difficult to estimate, 
a growing body of research has achieved success in 
determining whether a given network admits, for instance, bistability or oscillations, and even, in some cases, obtaining a mathematical description of
the parameter regions that give rise to such properties.
%As the corresponding
%  reaction networks can be very large and parameter values 
%are
%  difficult to obtain (if at all) a growing body of research tries to
%  answer the question if a particular mechanism can admit such
 % a qualitative property for some values of the parameters and, if so,
%  with obtaining mathematical descriptions of the corresponding
%  parameter regions. 
The aim of this Perspective is to highlight these recent mathematical developments.
% and then to demonstrate for specific PTM networks a flavor of the types of results we can now obtain.
% that can help analyze
% and  understanding
% PTM networks,
%a  flavor of the kind of
%  results that are obtainable.
% We also present a range of known results
%  for specific PTM networks to give a flavor of the kind of
%  results that are obtainable.

The basic building block of a PTM network is formed by the following two sets of reactions:
  \begin{equation}
    \label{eq:basic_network}
    S + M \lra S M \to S^* + M   \quad \text{ and }  \quad  S^*+U \lra S^* U \to S + U   ~,
  \end{equation}
in which a substrate $S$ forms a complex
%\footnote{
%      We use the term complex in the way it is used in chemistry. 
%In Chemical Reaction Network Theory, in contrast, the term refers to
%{\em any} left-hand or right-hand side of a reaction arrow (e.g., 
%has
 %     different meaning: it is used to denote the objects on both sides
  %    of the reaction arrows, for example 
%$S+M$, $S M$, $S^* + M$, and so on in~\eqref{eq:basic_network}).
%      form the left example above.}
  with a modifier $M$, and then $M$ and the modified substrate~$S^*$ 
%$S$ 
dissociate 
%the
%  modified substrate $S^*$ is released together with the $M$
  (left); this modification can be undone by another modifier $U$
  (right). 
  A PTM network is adapted from Eq.~\ref{eq:basic_network}
  by adding known biological interactions: for instance, the complexes $S M
  $ and $S^* U$ might undergo more modifications, there
  might be many substrates and modifiers, and a (modified)
  substrate of one reaction might be a modifier of another. 
%  PTM networks can be arbitrarily complicated: t

%\dblue{
%  Intermediate complexes like $S M$ and $S^* U$ are sometimes
%  neglected, but one often requires that in a PTM network, every
%  modifier has its antagonist (a single antagonist is allowed to undo
%  the modifications of several modifiers). This prevents an
%  accumulation of one of the modified substrates and seems to be the
%  reason for the rich mathematical structure of the underlying
%  equations. \marginpar{I am convinced that this is the case, but the
%    statement is very weak. Hence we might also drop the paragraph.}
%}

Arguably the most detailed and accurate description of the dynamics
of a PTM network is achieved at the level of mass-action. 
Here we focus on  systems in which the  spatial distribution of
concentrations can be neglected
  and where models therefore take the form of ordinary differential
  equations. Most models in the literature are of this type. 
%  even though spatial effects can influence the dynamics
  (When spatial effects may not be neglected  \cite{Kholodenko2006,KHOLODENKO901,Kholodenko2010}, partial differential equations
  must be considered and
%For systems where this does {\em not} hold, 
the connection between network structure and qualitative properties
has not yet been studied to the same extent; for some recent results,
see \cite{BifTheo-020,CasianMethodLines,diff-001}.)
  
  Even if spatial effects are neglected, the resulting dynamical
  models (systems of ODEs) 
%The
% resulting dynamical models in the form of ordinary differential
% equations,
% however, 
contain many variables and
  parameters  and are therefore difficult to analyze.
  Biophysicists and biochemists have long developed tools to deal with
  this situation, in particular the King-Altman method and the
  Michaelis-Menten reduction under the quasi-steady-state assumption
  proposed by Briggs and Haldane (see 
  \cite[Chapters~2\&4]{cornish2012fundamentals} and the references
  therein). 
  These methods guide the systematic elimination of
  variables and thus reduce the dimension of a
  system. 
%From a mathematical point of view it seems worthwhile to
%  mention that
It is important to note that the steady states of these reduced systems are in one-to-one
  correspondence with those of the original system \cite{feliu-wiuf-crn},
  while this need not be the case for the time-dependent trajectories of
  the respective ODE systems. 

In recent years,
  the abundance of PTM networks in biology and the accompanying
  need to analyze large networks has stimulated further research on
  the systematic elimination of variables. Much of this research
  originated at the interface of Chemical Reaction Network Theory,
  mathematical biology, and algebraic geometry. 
%We view some of the
 Some of the results obtained thereby are a natural extension of the 
King-Altman method and to some extent the Michaelis-Menten reduction. 
%While
 % some of the results are for general reaction networks, many are
 % specific to PTM networks. 
 In Section~3, we will summarize
%  some of
 the most consequential such elimination results for PTM
  networks %at the mass-action level 
(after introducing the mathematical
  formalism used to study PTM networks in Section~2). % and provide references to others.

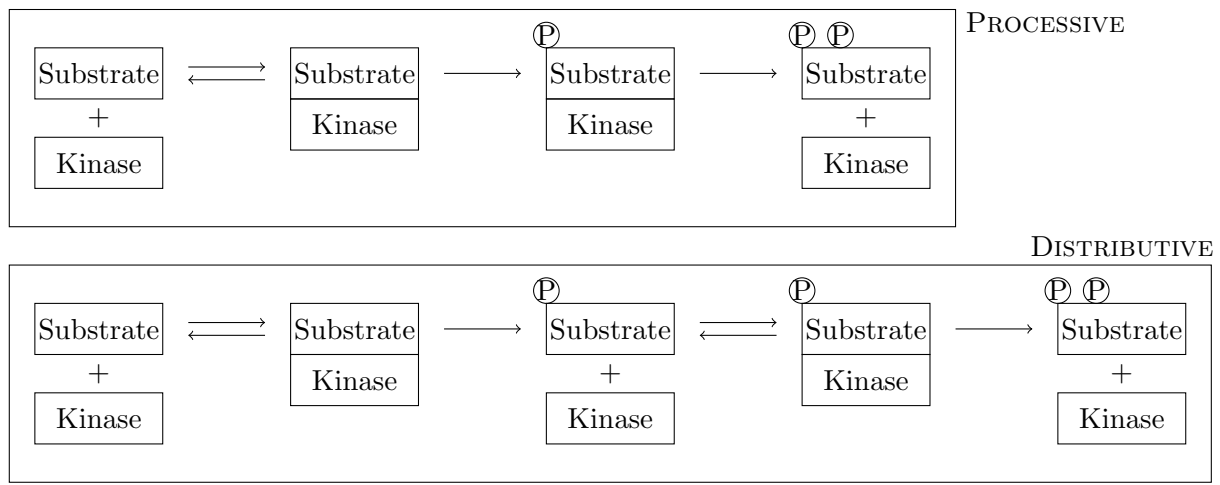
\begin{figure}[hbt]
\begin{center}
	\begin{tikzpicture}[scale=1.7]
%-----------------------------
% Processive
%-----------------------------
% BOX
\draw (0,3) rectangle (1, 3.4);
\draw (2,3) rectangle (3, 3.4);
\draw (4,3) rectangle (5, 3.4);
\draw (6,3) rectangle (7, 3.4);
% Box - kinase
\draw (0,2.3) rectangle (1, 2.7);
\draw (2,2.6) rectangle (3, 3);
\draw (4,2.6) rectangle (5, 3);
\draw (6,2.3) rectangle (7, 2.7);
% Phosphate groups
\draw (4,3.5) circle (0.1);	
\draw (6,3.5) circle (0.1);	
\draw (6.3,3.5) circle (0.1);	
% LABEL
    	\node[] at (0.5, 3.2) {Substrate};
    	\node[] at (2.5, 3.2) {Substrate};
    	\node[] at (4.5, 3.2) {Substrate};
    	\node[] at (6.5, 3.2) {Substrate};
    	\node[] at (0.5, 2.85) {$+$};
    	\node[] at (0.5, 2.5) {Kinase};
    	\node[] at (2.5, 2.8) {Kinase};
    	\node[] at (4.5, 2.8) {Kinase};
    	\node[] at (6.5, 2.85) {$+$};
    	\node[] at (6.5, 2.5) {Kinase};
% LABEL - P's
    	\node[] at (4, 3.5) {P};
    	\node[] at (6, 3.5) {P};
    	\node[] at (6.3, 3.5) {P};
% ARROWS
	 \draw[->] (1.2, 3.25) -- (1.8, 3.25);	
	 \draw[<-] (1.2, 3.15) -- (1.8, 3.15);	
	 \draw[->] (3.2, 3.2) -- (3.8, 3.2);	
	 \draw[->] (5.2, 3.2) -- (5.8, 3.2);
% BOX
\draw (-0.2,3.7) rectangle (7.2,2);
% LABEL
\node[] at (7.9, 3.6) {{\sc Processive}};
%-----------------------------
% Distributive
%-----------------------------
% LABEL
\node[] at (8.5, 1.85) {{\sc Distributive}};
% BOX
\draw (0,1) rectangle (1, 1.4);
\draw (2,1) rectangle (3, 1.4);
\draw (4,1) rectangle (5, 1.4);
\draw (6,1) rectangle (7, 1.4);
\draw (8,1) rectangle (9, 1.4);
% Box - kinase
\draw (0,0.3) rectangle (1, 0.7);
\draw (2,0.6) rectangle (3, 1);
\draw (4,0.3) rectangle (5, 0.7);
\draw (6,0.6) rectangle (7, 1);
\draw (8,0.3) rectangle (9, 0.7);
% LABEL
    	\node[] at (0.5, 1.2) {Substrate};
    	\node[] at (2.5, 1.2) {Substrate};
    	\node[] at (4.5, 1.2) {Substrate};
    	\node[] at (6.5, 1.2) {Substrate};
    	\node[] at (0.5, 0.85) {$+$};
    	\node[] at (0.5, 0.5) {Kinase};
    	\node[] at (8.5, 1.2) {Substrate};
    	\node[] at (8.5, 0.85) {$+$};
    	\node[] at (8.5, 0.5) {Kinase};
    	\node[] at (2.5, 0.8) {Kinase};
    	\node[] at (4.5, 0.85) {$+$};
    	\node[] at (4.5, 0.5) {Kinase};
    	\node[] at (6.5, 0.8) {Kinase};
% Phosphate groups
\draw (4,1.5) circle (0.1);	
\draw (6,1.5) circle (0.1);	
\draw (8,1.5) circle (0.1);	
\draw (8.3,1.5) circle (0.1);	
% LABEL - P's
    	\node[] at (4, 1.5) {P};
    	\node[] at (6, 1.5) {P};
    	\node[] at (8, 1.5) {P};
    	\node[] at (8.3, 1.5) {P};	
% ARROWS
	 \draw[->] (1.2, 1.25) -- (1.8, 1.25);	
	 \draw[<-] (1.2, 1.15) -- (1.8, 1.15);	
	 \draw[->] (3.2, 1.2) -- (3.8, 1.2);	
	 \draw[->] (5.2, 1.25) -- (5.8, 1.25);	
	 \draw[<-] (5.2, 1.15) -- (5.8, 1.15);	
	 \draw[->] (7.2, 1.2) -- (7.8, 1.2);
% BOX
\draw (-0.2,1.7) rectangle (9.2,0);
    	\end{tikzpicture}
\end{center}
\caption{Dual-site phosphorylation may be {\em processive} (top) or {\em distributive} (bottom) or somewhere in between; cf.\ \cite[Figure 2B]{salazar}.}
\label{fig:proc-dist}
\end{figure}

%  \marginpar{We could also state what is known about n-site systems.}
  In Section~4 we will turn to a specific class of PTM networks, those
  describing the multisite phosphorylation/dephosphorylation of a single substrate.
% and a
%  subclass describing double phosphorylation. 
%These systems have been studied in a variety of publications and a lot is
Much has been discovered in recent years about 
%  known about 
the steady states and the dynamics of these systems --
  more than for general PTM networks. 
What has been revealed is a close connection between a network's structure and its qualitative dynamics.  Consider, for instance, 
  the sequential, double phosphorylation (also called dual-site phosphorylation) of a protein 
in which the enzymatic mechanism is 
fully {\em processive} versus those with fully {\em distributive} mechanism (see Figure~\ref{fig:proc-dist}). 
 Processive systems are globally stable -- convergent to a unique steady state -- while 
 distributive systems may be bistable.  
These properties -- and those of several related networks -- are summarized later in Figure~\ref{fig:multisite-aims}.

%-----------------------------
% DIST. VS. PROC. FIGURE
%-----------------------------

In this Perspective, and especially at the close, we will highlight important future directions.

 \subsection{Related review articles}
%\dblue{
%  This section seems still too informal, we need all the references,
%  but we could/should mention that there is currently no review that
%  summarizes the work on elimination (Feliu, Wiuf, Gunawardena) and on
%  the multisite phosphorylation of a single enzyme. In this context we
%  could also cite Katharina's work on distributive
%  phosphorylation. Maybe call Section 4/5 multisite phosphorylation of
%  a single enzyme with subsections purely distributive, purely
%  processive, mixed (mixed being a section containing questions for
%  future research). And maybe a section on double phosphorylation
%  where we could have Fig.~\ref{fig:multisite-aims}.
%}
Excellent reviews on various aspects of PTM networks
%multisite phosphorylation systems (the best-studied PTM systems) 
have been published (which we highlight below).  Our goals therefore are complementary to what earlier reviews accomplished and, to some extent, pick up where they left off.  Indeed, there is currently no review summarizing recent progress on elimination for PTM networks or the dynamics of
 multisite phosphorylation networks involving a single substrate.

As mentioned earlier, we are interested in the question of how PTM networks process information.  
Theoretical aspects of this question, and experimental approaches also, were reviewed
by Prabakaran {\em et al.}~\cite{escape}.  
Our Perspective, in contrast, is limited to the theory side, and specifically to the 
question of how a PTM network constrains the resulting dynamics (which in turn constrain the capacity for information encoding).

Another related review is that of Salazar and H\"ofer~\cite{salazar}, which describes 
the many biological functions of (multisite) phosphorylation (the best-studied PTM), a thorough list of possible mechanisms, and the resulting kinetic models.  
%, we refer the reader to the 2009 review by Salazar and H\"ofer~\cite{salazar}.  
Their review, % on multisite phosphorylation systems, 
to abridge its title, goes ``from molecular mechanisms to kinetic models'' --
whereas, here we go from kinetic models to dynamics.  
% To be clear,
% Salazar and H\"ofer did examine
% dynamics (including ultrasensitivity
% and bistability but not
% oscillations), but our scope is
% narrower (we examine only certain
% mechanisms) and we focus on what we
% have learned about these kinetic
% models since their review came out.
  To be clear, Salazar and H\"ofer did examine dynamics (including
  ultrasensitivity and bistability but not oscillations), but not for all
  the systems described here. 

Finally, much of the interest in phosphorylation networks is due to their role in MAPK cascades.  The dynamics of MAPK cascades was reviewed by Hell and Rendall in~\cite{hell-2015}.

%Another review article:~\cite{models}.  

%---------------------------------
%: BACKGROUND
%---------------------------------
\section{Background} \label{sec:background}
{\em Post-translational modification} % (PTM) 
is a biochemical alteration of a protein that occurs after its mRNA has been translated into a sequence of amino acids.
The most common PTM is {\em phosphorylation}, the enzyme-mediated addition of a phosphate group to a protein substrate, which, to quote from~\cite{25}, ``can alter its behavior in almost every conceivable way.''
%``The phosphorylation of a protein can alter its behavior in almost every conceivable way.''~\cite{25} (abstract)
  The basic phosphorylation/dephosphorylation mechanism is a relabeled version of the building block in Eq.~\ref{eq:basic_network}:
%variant of (\ref{eq:basic_network})
%  obtained by relabelling:
  \begin{equation}
    \label{eq:basic_phospho}
    S_0 + K \lra S_0 K \to S_1+K \quad \text{ and }  \quad S_1 + F \lra S_1 F \to S_0+F~,
  \end{equation}
  where a {\em kinase} and a {\em phosphatase} enzyme ($K$ and $F$) act as
  modifiers, % $M$ and $U$, 
the non-phosphorylated (unmodified) protein is
  denoted by $S_0$, and the phosphorylated (modified) protein is $S_1$. 
% The basic mechanism is: $S_0 + K \lra S_0 K \to S_1+K$, where $K$ is the kinase enzyme and $S_0$ and $S_1$ represent the substrate with, respectively, 0 and 1 phosphate groups attached.
The mechanism in Eq.~\ref{eq:basic_phospho} also describes other PTMs, such as acetylation, the
addition of an acetyl group.  Although some PTMs are non-binary (more
than one modification is possible, even at a single
site~\cite{escape}), we will focus on PTMs built on %this simple
the mechanism in Eq.~\ref{eq:basic_phospho} and use, for ease, %simplicity, 
the language of
phosphorylation. %restrict our attention to phosphorylation. 

So far we have discussed phosphorylation on a single site.  
Many substrates, however, have more than one site at which phosphate
groups can be attached 
(in which case we continue to use subscripts to denote the number of phosphorylations, 
e.g., $S_2$ is a doubly phosphorylated protein).
%, and $S_n$ is a protein that has $n$ phosphate groups attached).
%\dblue{(
 % in this case the number of phosphorylations is denoted by a subscript, e.g.\ $S_2$
 % for a double phosphorylated protein and $S_n$ for a protein that is
 % phosphorylated $n$-times).}
Such multisite phosphorylation %-- 
%or {\em dephosphorylation} which removes phosphate groups -- 
may be 
  {\em sequential} or {\em random}, or somewhere in between, 
and {\em processive} or {\em distributive}, or somewhere in
between~\cite{Guna_threshold, PM, salazar}. 
  In {\em sequential} phosphorylation
  the last phosphate group to be attached is the first to be removed, so,
  for example, 
  there is a unique once-phosphorylated substrate ($S_1$). This Perspective focuses solely on sequential phosphorylation. 
(In {\em random} phosphorylation, phosphate groups can be added in any order.)
{\em Processive} phosphorylation, as depicted in Figure~\ref{fig:proc-dist},
requires an enzyme and 
substrate to bind only once in order to add 
  or remove 
all phosphate groups.
{\em Distributive} phosphorylation, in contrast, requires multiple bindings
to add 
  or remove 
all phosphate groups, because 
   at most one group is added with every binding of enzyme
  and protein. 
%For an example see Example~\ref{ex:mixed} below, where
 % phosphorylation is processive and dephosphorylation is
 % distributive. 

\begin{figure}[hb]
\begin{center}
	\begin{tikzpicture}[scale=1]
% LABEL: A to E
    	\node[above left] at (0,0.25) {(E)};
    	\node[above left] at (0,3.25) {(A)};
    	\node[above left] at (0,1.75) {(C)};
    	\node[above right] at (15,1.75) {(D)};
    	\node[above right] at (15,3.25) {(B)};
%    	\draw[help lines] (0,0) grid (10,5);
	\draw (0,0) rectangle (6.5,1.2);
    	\node[above] at (3.2,0.5) {Mixed-mechanism dual-site};
    	\node[above] at (3.2,0) {(unique steady state, oscillations)}; %, Problem \ref{q:osc})};
	\draw (0,1.5) rectangle (6.5,2.7);
    	\node[above] at (3.2,2) {ERK-mechanism dual-site};
    	\node[above] at (3.2,1.5) {(bistable, oscillations)};
	\draw (0,3) rectangle (6.5,4.2);
    	\node[above] at (3.2,3.5) {Fully {distributive} dual-site};
    	\node[above] at (3.2,3) {(bistable)};%, Hopf bifurcation reported)};
% the ones they limit to...
	\draw (10,1.5) rectangle (15,2.7);
    	\node[above] at (12.5,2) {Fully {processive} dual-site};
    	\node[above] at (12.5,1.5) {(globally stable)};
	\draw (10,3) rectangle (15,4.2);
    	\node[above] at (12.5,3.5) {Michaelis-Menten dual-site};
    	\node[above] at (12.5,3) {(bistable, no oscillations)};
% limiting "phrase" to indicate...
    	\node[above] at (8.3,3.2) {{\em \dots reduces to \dots}};
    	\node[above] at (8.3,2) {{\em \dots reduces to \dots}};
%    	\node[above] at (8.3,1.5) {(Problem \ref{q:level-proc})};
 	\end{tikzpicture}
\end{center}
\caption{Dual-site phosphorylation/dephosphorylation networks, also called {\em futile cycles}: five types and their properties. A network reduces to another if the latter approximates the former if certain conditions on the parameters hold.  See \S \ref{sec:a}--\S \ref{sec:e}.} 
\label{fig:multisite-aims}
\end{figure}
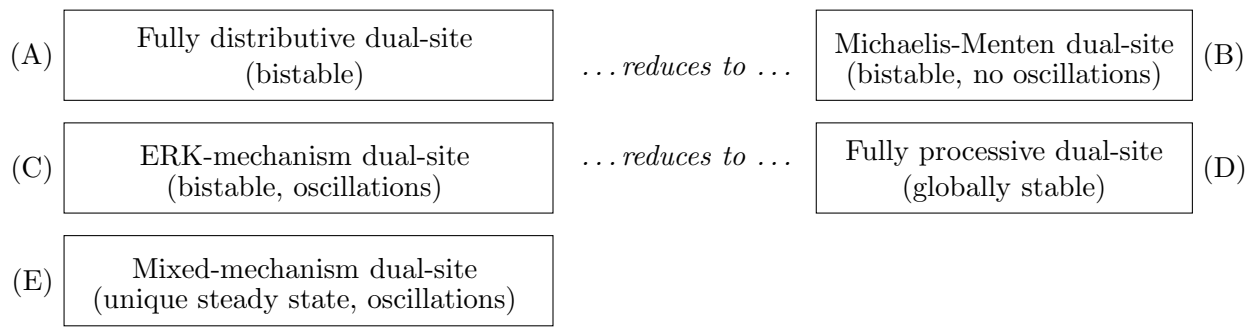

\begin{example}[Mixed-mechanism network] \label{ex:mixed}
The following is a {\em mixed-mechanism} dual-site network (of Figure~\ref{fig:multisite-aims}(E)), in which phosphorylation is processive and dephosphorylation is distributive~\cite{SK}:
\begin{equation}
  \label{eq:mixed-network}
  \begin{split}
  \xymatrix{
    % PROCESSIVE
    S_0 + K  \ar @<.4ex> @{-^>} [r] ^-{k_1}
    &\ar @{-^>} [l] ^-{k_{2}} 
    S_0 K \ar 
    % @<.4ex> @{-^>} 
    [r] ^-{k_3} 
    & S_1 K \ar %@<.4ex> @{-^>} 
    [r] ^-{k_4}
    & S_2 + K\\
    % DISTRIBUTIVE
    S_2 + F \ar @<.4ex> @{-^>} [r] ^-{k_{5}}
    &\ar @{-^>} [l] ^-{k_6} S_{2} F 
    \ar %@<.4ex> @{-^>} 
    [r] ^-{{k_7}} 
    % &\ar @{-^>} [l] ^-{\ell_{2n-2}} \hdots \ar @<.4ex> @{-^>} [r]
    % ^-{\ell_5}
    % &\ar @{-^>} [l] ^-{\ell_4} S_2 F \ar @<.4ex> @{-^>} [r]
    % ^-{\ell_3}
    % &\ar @{-^>} [l] ^-{\ell_2} S_1 F \ar [r]
    % ^-{\ell_1}
    &S_1 + F
    \ar @<.4ex> @{-^>} [r] ^-{k_{8}}
    &\ar @{-^>} [l] ^-{k_9}
    S_{1} F \ar %@<.4ex> @{-^>} 
    [r]^-{{k_{10}}} & S_0 + F
  }
  \end{split}
\end{equation}

%Here, $S_i$ denotes a substrate with $i$ phosphate groups attached, and $K$ and $F$ are, respectively, a {\em kinase} and a {\em 
%phosphatase} enzyme.  
% The notation $S_i$ reflects the fact that this network is {\em sequential}: the last phosphate group to be attached is the first to be removed, so there is a unique %singly
% once-phosphorylated substrate, and we denote it by $S_1$.

Under the commonly used assumption that the (cellular) volume is constant, 
the network in Eq.~\ref{eq:mixed-network} gives rise, via mass-action kinetics,
to the following ODEs:
%The mixed-mechanism phosphorylation system in Figure \ref{fig:multisite-aims}(E):
\begin{align} 
    \notag
 \dot x_1 &~=~     -k_1 x_1 x_2+k_2 x_3+k_{10} x_9 \\
    \notag
 \dot x_2 &~=~      -k_1 x_1 x_2+k_2 x_3+k_4 x_4 \\
    \notag
 \dot x_3 &~=~      k_1 x_1 x_2-(k_2+k_3) x_3  \\
%\end{align}
%\begin{align}
    \notag
 \dot x_4 &~=~      k_3 x_3- k_4 x_4 \\
    \label{eq:OEs-mixed}
 \dot x_5 &~=~      k_4 x_4-k_5 x_5 x_6+k_6 x_7 \\
    \notag
 \dot x_6 &~=~      -k_5 x_5 x_6-k_8 x_8 x_6+(k_6+k_7) x_7+(k_9+k_{10}) x_9 \\
    \notag
 \dot x_7 &~=~      k_5 x_5 x_6-(k_6+k_7) x_7 \\
    \notag
 \dot x_8 &~=~      k_7 x_7-k_8 x_6 x_8+k_9 x_9 \\
    \notag
 \dot x_9 &~=~      k_8 x_6 x_8-(k_9+k_{10}) x_9~,
\end{align}
where $x_1,x_2, \ldots, x_9$ denote, respectively, the concentrations of the species
$S_0$, $K$, $S_0K$, $S_1 K$, $S_2$, $F$, $S_2 F$, $S_1$, $S_1 F$.  The total amounts of free and bound enzyme or substrate, denoted by 
$(\Ktot, \Ptot, \Stot )\in \R_{>0}^3$,
remain constant as the dynamical system in Eq.~\ref{eq:OEs-mixed} progresses, so we obtain the following {\em conservation laws}:
\begin{align}
\label{eqn:conservation}
\Ktot ~=~ x_2 + x_3 + x_4~, ~ 
\Ptot ~=~ x_6+x_7+x_9  ~, ~
\Stot ~=~ x_1+x_3+x_4+x_5+x_7+x_8+x_9  ~.
\end{align}
Thus, each trajectory $x(t)$ of Eq.~\ref{eq:OEs-mixed} is confined to
an {\em invariant set} defined by some $(\Ktot, \Ptot, \Stot )$: 
%\dblue{
%  should we add: invariant set is defined by the initial value, the
%  fact that solutions are constrained to the set is important
%  in discussing the number of steady states? Maybe look ahead and say
%  that whenever elimination works as described in
%  Section~\ref{sec:intermediates} one obtains infinitely many steady
%  states but that these intersect a given invariant set only in
%  finitely many points (in general). However, form an algebraic point
%  of view, it is unclear why this is the case, precise conditions on
%  the network guaranteeing this seem unknown.
%}
\begin{align} \label{eq:invt-set}
\{x \in \mathbb{R}_{\geq 0}^9 \mid 
    \text{ the conservation equations $\eqref{eqn:conservation}$ hold} \}~.
\end{align}
\end{example}

One focus of this Perspective is on the dynamics arising from 
  multisite phosphorylation networks such as the mixed-mechanism
  network of Example~\ref{ex:mixed} and those displayed in
  Figure~\ref{fig:multisite-aims}. We report what is known about
  their qualitative dynamics: we describe, whether or not a network
  is {\em bistable}, i,e. whether or not it has the capacity for two
  or more steady states, stable to small perturbations, within some
  invariant set in Eq.~\ref{eq:invt-set}.
% the mixed-mechanism network (Example~\ref{ex:mixed}) and others
% (such as those displayed in Figure~\ref{fig:multisite-aims}). Is
% the network {\em bistable}, i.e., does it have the capacity for two or
% more steady states, stable to small perturbations, within some
% invariant set in Eq.~\ref{eq:invt-set}?
(The biological significance of bistability has been described, for
instance, in \cite{thomson2009unlimited}). If not, is the network {\em
  globally stable}, with a unique steady state in each invariant set to which
all trajectories converge? Or, does the network admit {\em
  oscillations}, i.e., periodic orbits?
To answer such questions, one key technique is elimination, which we discuss next.

%---------------------------
% \begin{figure}
%     \centering
%     \includegraphics{figure2_ex_2dd_4p}
%     \caption{\label{fig:examplel} A PTM network. of four modifications. (a) interconnection of modules from Fig.~\ref{fig:modules}. (b) network at the mass action level.} 
% \end{figure}
%---------------------------
% \begin{figure}
%     \centering
%     \includegraphics[width=0.6\linewidth,angle=-90]{figure3_ss_para_2}
%     \caption{\label{fig:odes_para} ODES (a), equations defining $k_M$- and $k_{cat}$-values (b) and (monomial) parametrization of positive steady states (c).}
% \end{figure}
%---------------------------

%---------------------------------
%: INTERMEDIATES
%---------------------------------
\section{Elimination and the role of intermediate complexes} \label{sec:intermediates}
We begin this section with the well-known reversible Michaelis-Menten mechanism
  (cf.\ \cite{cornish2012fundamentals}):\\
  \begin{minipage}[t]{0.4\linewidth}
    \underline{Reaction Network}
    \begin{align*}
      &\xymatrix{
        S_{0}+ K \ar @<.4ex> @{-^>} [r] ^-{k_1}
      &\ar @{-^>} [l] ^-{k_2}  S_{0}K \ar @<.4ex> @{-^>} [r] ^-{k_3}
      &\ar @{-^>} [l] ^-{k_4}   S_{1}+K 
        }\\
      \intertext{\underline{Conservation laws}}
      & x_1 + x_3 + x_4 = \Stot  \\ %s_0 \\
      & x_2 + x_3 = \Ktot %e_0
    \end{align*}
  \end{minipage}
  \begin{minipage}[t]{0.55\linewidth}
    \underline{ODEs (using $x_1$ for $S_0$, $x_2$ for $K$, $x_3$ for
      $S_0K$, and $x_4$ for $S_1$):}
    \begin{align*}
      \dot x_1 &= -k_{1} x_{1} x_{2}+k_{2} x_{3} \\
      \dot x_2 &= -k_{1} x_{1} x_{2}+(k_{2} +k_{3}) x_{3}-k_{4}
                 x_{2} x_{4} \\ 
      \dot x_3 &=  \phantom{-}k_{1} x_{1} x_{2}-(k_{2}+k_{3}) x_{3}+k_{4} x_{2}
                 x_{4} \\
      \dot x_4 &= \phantom{-}k_{3} x_{3}-k_{4} x_{2} x_{4}
    \end{align*}
  \end{minipage}\\
 % Often %one is in the situation that
The total amount of substrate $\Stot$ is
often much %  significantly 
larger than %the total amount of 
that of the enzyme $\Ktot$
  (Briggs-Haldane assumption \cite{cornish2012fundamentals}), and one can
  therefore assume $\dot x_3 \approx 0$. It is then standard practice
  to use the equation $\dot x_3 = 0$ % for $\dot x_3$ 
and the %conservation
 relation $ x_2 + x_3 = \Ktot $
%of
 % the enzyme
 to \emph{eliminate}  the variables $x_2$ and $x_3$.
% from the system. 
That is, one solves for $x_2$ and $x_3$ in terms
  of $x_1$, $x_4$, $\Ktot$, and the rate constants (parameters): 
%\marginpar{mention: after relabeling the classical form found in
 %   textbooks?} 
  \begin{align} \label{eq:param-mm}
    x_2 &= \frac{\Ktot (k_{2}+k_{3})}{k_{2}+k_{3}+k_{1} x_{1}+k_{4} x_{4}}, &
    x_3 &=\frac{\Ktot (k_{1} x_{1}+k_{4}x_{4})}{k_{2}+k_{3}+k_{1}
          x_{1}+k_{4} x_{4}}~, \\
    \intertext{and substitutes these expressions into %the equations for 
    $\dot x_1$ and
    $\dot x_4$ to obtain the ODEs of the 
    %\lq reduced\rq{} 
    reduced
    system:}
    \dot x_1 &= \frac{\Ktot (-k_{1} k_{3} x_{1}+k_{2} k_{4}
               x_{4})}{k_{2}+k_{3}+k_{1} x_{1}+k_{4} x_{4}} = -\dot x_4~. \notag
  \end{align}%
%  From an algebraic geometry point of view we have obtained 
Mathematically, Eq.~\ref{eq:param-mm} is a {\em parametrization} of $x_2$ and $x_3$ in terms of $x_1$, $x_4$, $\Ktot$,
  and rate constants. % Note that the parametrization of $x_2$ and
%   $x_3$ corresponds to the steady state values of the original system and
%   that by setting either one of the ODEs for $\dot x_1$ or $\dot x_4$
%   to zero one can obtain a parametrization of all steady states (where
%   $x_2$, $x_3$ and either $x_1$ or $x_4$ are given in terms of $e_0$,
%   the rate constants and $x_4$ or $x_1$). 
% }
%

 % Of course 
One can apply similar reasoning to PTM networks, removing
intermediate complexes like $S_0K$ to obtain a reduced system with fewer variables. 
Although this process can 
become arduous for large networks, the well-known King-Altman method
provides a systematic way to eliminate intermediates
\cite{cornish2012fundamentals}.
However, this method
relies at least implicitly on the
  ability to identify the enzyme (to guide the choice of
  variables to solve for). Furthermore, it does not explain the
  positivity of the parametrization. The first requirement, in
  particular, limits the applicability of the method as
 it is, frequently the case (for
  example, in cellular signaling) that the
  substrate of one modifier acts as a modifier for a second
  substrate. Such cascades cannot be handled by the King-Altman
  method. 

Fortunately, both concerns were resolved recently.
First, positivity in the King-Altman framework 
was explained
  by  Thomson and Gunawardena  \cite{TG}, who initiated the theory underlying elimination in PTM networks (see also~\cite{linear-framework}). 
Subsequently, Feliu and Wiuf extended this to address the second concern: how to eliminate 
in cascades~\cite{feliu-signaling} and other general networks~\cite{feliu-wiuf-crn,
feliu-wiuf-ptm} (see also \cite{saez-gph}).
%, and then, with Saez, obtained graphical criteria for when such elimination is feasible \cite{saez-gph}.  
These works describe procedures to systematically eliminate
  intermediates and some substrates.  
In fact, the resulting parametrizations arise from spanning
  trees of certain graphs -- much like King-Altman!
 
%As for cascades and other general PTM
%  networks,
%%   and for more general PTM
%%  networks, including cascades, by 
%Feliu and Wiuf \cite{feliu-wiuf-crn,feliu-wiuf-ptm} 
%%These references 
%  describe procedures to systematically eliminate
%  intermediates and some of the substrates. 
  
%Let us say more about the parametrizations of Feliu and Wiuf.
%  First, they distinguished between substrates (which can be modified)
%and intermediates (including enzyme-substrate complexes).
% The above example, for instance, consists of three
%  substrates ($S_0$, $S_1$, and $K$) and one intermediate
%  ($S_0K$). 
%Next, they gave conditions that guarantee 
%that every variable representing an intermediate
%and some of those representing substrates can be parameterized at steady state in
%  terms of a subset of the substrate
%  variables, %These variables are called 
%  called the {\em core variables}. 
%  In our example, $x_1$ and $x_4$ are %thus
%  core variables. They showed, moreover, that this parametrization 
%  is given by rational functions (i.e., the fraction of
%  two polynomials) with positive denominators 
%  -- as in the %earlier 
%  parametrization of $x_2$ and $x_3$ in (\ref{eq:param-mm}). 

\subsection{Elimination and steady-state parametrizations}
%%%%%
  For the Michaelis-Menten mechanism above, elimination
  was guided by biophysical insight as well as the aim to simplify the
  system and understand the dynamics (of the simplified
  system). If one were interested only in the positive steady states
  of the original ODEs, or in the number of steady states, 
% for given values  of $K_{tot}$ and $S_{tot}$, 
then one might pursue a different approach which also uses a parametrization.  
Specifically, we set to zero the right-hand sides of the differential equations, and then 
solve for each $x_i$ in terms of just $x_2$ and $x_4$, % (for instance, using {\tt Mathematica}) 
and see that the set of all steady states is the 2-dimensional image of the map
 % $\psi = \psi_{k_1, \dots, k_{4} } : 
%$\psi:
$ \mathbb{R}^2_{>0}  \to \mathbb{R}^{4}_{>0} $ given by:  
  \begin{align} \label{eq:param}
    (x_2, x_4) ~\mapsto ~ \left( \frac{k_{2} k_{4} }{k_{1} k_{3}} x_{4},~ x_2, ~
    \frac{k_{4} }{k_{3}} x_{2} x_{4},~ x_4\right)~.
  \end{align}  
%This function is a {\em steady-state parametrization}.
%For example, first solve the steady state equations for
%  $x_1$ and $x_3$ in terms of $x_2$, $x_4$ and the rate constants to obtain
%  \begin{displaymath}
%    x_{1} = \frac{k_{2} k_{4} x_{4}}{k_{1} k_{3}} \text{ and } 
%    x_{3} = \frac{k_{4} x_{2} x_{4}}{k_{3}}
%  \end{displaymath}
%  and then determine the intersection(s) with the invariant
%  sets. Note that this is  equivalent to eliminating $x_1$ and
%  $x_3$. In particular, we have obtained a parameterization of
%  \emph{all} positive solutions to the steady state equations:
%  \begin{equation}
%    \label{eq:para_mm}
%    x^*(x_2,x_4) = \left( \frac{k_{2} k_{4} x_{4}}{k_{1} k_{3}}, x_2,
%    \frac{k_{4} x_{2} x_{4}}{k_{3}}, x_4\right).
%  \end{equation}
%  For any positive choice of $x_2$, $x_4$ and the rate constants, the
%  corresponding $x^*$ will be a (positive) steady state. The
This {\em steady-state parametrization} is particularly simple, as it
  consists of the \emph{monomials} $x_2$, $x_4$, and $x_2x_4$ (or their scalar multiples). It is
  consequently called a \emph{monomial parametrization}.

 % As a consequence of the results of Feliu and Wiuf 
The existence of a positive parametrization is guaranteed for PTM networks
  \cite{feliu-wiuf-ptm,messi,TG}. 
  %It can be computed by  eliminating
 % the intermediate complexes (i.e., by computing spanning trees). 
  However, such a parametrization need not be monomial.

As for the number of steady states, this corresponds mathematically to 
counting the number of times the image of the parametrization intersects some invariant set. 
If at least one such intersection (for some choice of rate constants) has at least two elements (i.e., at least two positive steady states in some invariant set), then we say that the network is {\em multistationary}.  Deciding whether a network is multistationary is in general difficult~\cite{mss-review}, 
but this can be accomplished easily when the parametrization is monomial
  \cite{signs,TSS}. Such an argument, for instance, was used to prove
  that 
  a reversible version of
the following {\em fully processive, $n$-site phosphorylation network} (for which the $n=2$ case is the network of Figure~\ref{fig:multisite-aims}(D)) is {\em not} multistationary~\cite{ConradiShiu}:
\begin{equation} % Make first reversible.
\label{eq:n-site-processive}
\begin{split}
  \xymatrix{
    S_0 + K  \ar @<.4ex> @{-^>} [r] ^-{}
    &\ar @{-^>} [l] ^-{} S_0 K \ar [r] ^-{}
    &S_1 K \ar [r] ^-{}
    &\hdots \ar  [r] ^-{}
    & S_{n-1} K \ar [r] 
    ^-{}
    & S_n + K\\
    S_n + F \ar @<.4ex> @{-^>} [r] ^-{}
    &\ar @{-^>} [l] ^-{} S_{n} F \ar  [r] ^-{} 
    ^-{} 
    &\hdots \ar [r]
    ^-{}
    &S_2 F \ar  [r]
    ^-{}
    &S_1 F \ar [r]
    ^-{}
    &S_0 + F
  }
\end{split}
\end{equation}
Another approach to precluding multistationarity in this network 
%via removing intermediates, 
is described below in Section~\ref{sec:remove-intermediates}.
  
%  While such a parameterization contains information about \emph{all}
%  positive steady states, one is often interested in the (number of)
%  steady states in a given invariant set. Mathematically this
%  corresponds to computing the intersection of the parameterization
%  and the invariant set. And if this intersection contains at least
%  two elements, then one speaks of
%  \emph{multistationarity}. Determining this intersection is in
%  general difficult, unless the parameterization is monomial
%  \cite{signs,TSS}; for an example see the application of the argument to 
%  (\ref{eq:n-site-processive}) is described in \cite{ConradiShiu}.

  For large (not necessarily PTM) networks  it can be challenging to obtain a
  parametrization, as there is (currently) no %formal
  procedure apart from computing spanning trees 
  (unless the network has more structure~\cite{messi}). 
  %In practice we
  %recommend using a computer algebra system like {\tt Mathematica} or 
  %{\tt Maple}. 
  Guided by
  experience, we suggest the following approach: if $p$ is the number of conservation laws, then we
  expect to parametrize the steady states in
  terms of $p$ variables (there is no guarantee though). Therefore, using a computer algebra system like {\tt Mathematica} or 
  {\tt Maple}, solve for all but $p$ variables, eliminating all
  intermediates (and some substrates)~\cite{feliu-wiuf-ptm,TG}. If the result is not
  satisfactory, try another subset of variables.

Finally, we can express a steady-state parametrization in terms of 
biochemically meaningful parameters, namely, 
the catalytic constants of the enzyme and the $K_m$ values.
%  With the help of a computer algebra system one might also
%  incorporate a parameter change. Suppose we want to express the
For instance, for the above parametrization (\ref{eq:param-mm}),
%in terms of the $K_m$ values and the catalytic constants of the enzyme. In this system
  there are two $K_m$ values: $K_{m,1} = \frac{k_2+k_3}{k_1}$ for the
  substrate $S_0$ and $K_{m,2} = \frac{k_2+k_3}{k_4}$ for the
  substrate $S_1$.  
%  We can introduce these by adding the equations
%  $k_1\cdot K_{m1}-(k_2+k_3)=0$ and $k_4 \cdot K_{m2}-(k2+k3) = 0$,
%  and then solving for $x_1$, $x_3$, $k_1$, and $k_2$.
We solve these $K_m$ equations for $k_1$ and $k_4$ and substitute these
expressions into the parametrization (\ref{eq:param-mm}), which becomes:
\begin{align*} % \label{eq:param-catalytic-constants}
    (x_2, x_4) ~\mapsto ~ \left(  \frac{k_2 K_{m,1}}{k_3 K_{m,2}} x_{4},~ x_2, ~
   \left(1+\frac{k_{2}}{k_{3}}\right) \frac{1}{K_{m,2}} x_{2} x_{4},~ x_4\right)~.
  \end{align*}  
Such an approach was used to analyze how catalytic constants enable the emergence of 
bistability in the fully distributive dual-site phosphorylation system (the network of Figure \ref{fig:multisite-aims}(A))~\cite{a6maya}. 

%  We can introduce these by adding the equations
%  \begin{displaymath}
%    k_1*K_{m1}-(k_2+k_3)=0 \text{ and } k_4*K_{m2}-(k2+k3) = 0
%  \end{displaymath}
%  and solve for $x_1$, $x_3$ and $k_1$, $k_2$. The result
%  \begin{displaymath}
%    x^*(x_2,x_4)=\left( \frac{k_2}{k_3}\frac{K_{m1} x_{4}}{K_{m2}}, x_2,
%      \left(1+\frac{k_{2}}{k_{3}}\right) \frac{x_{2} x_{4}}{K_{m2}}, x_4 \right)
%    \text{ and } k_{1} = \frac{k_{2}+k_{3}}{K_{m1}}, 
%    k_{4} = \frac{k_{2}+k_{3}}{K_{m2}}
%  \end{displaymath}
%  is a parameterization of the positive steady states in terms of the
%  $K_m$-values and the catalytic constants $k_2$ and $k_3$. 
%  Clearly ob can do this, whenever one obtains a parameterization with
%  a computer algebra system.

\subsection{Intermediates and steady states} \label{sec:remove-intermediates}
Going beyond parametrizations, Feliu and Wiuf also
  analyzed elimination at the level of the reaction network, and its effect on steady states and their stability. Here
%  iterative 
removal of intermediates yields a reduced
  reaction network, 
 %  that they call the {\em core network}
%, called  the core network in
  and they proved
  the following:
  % the following relationship between the steady states of these two networks:
% original network and those of the core network:
 if the reduced network admits at least $N\geq 1$ (locally stable) steady states (for
  some parameter values), then the original network also has at 
least $N$ (locally stable) steady states (for some parameters) \cite{FeliuWiuf}. 
(See~\cite{BP-inher,ME_entrapped,mss-review} for other results that ``lift'' multistationarity/bistability from small networks to large.)
%re exist parameter values such
%  that the original network has at least $N$ steady states. 
%In the aforementioned reference the relation
 % between steady states in the complete network and in the core
 % network is described: if the core network has $N\geq 1$ steady states (for
 % some parameter values) then there exist parameter values such
 % that the original network has at least $N$ steady states. 
% The standard reaction network underlying single-site phosphorylation, 
% as we saw above,
% is $S_{0}+ K \lra S_{0}K \to  S_{1}+K$ (where $S_i$ is the substrate with $i$ phosphate groups attached, and $K$ is the kinase). However, perhaps $S_{0}+K  \lra S_{0}K \to { S_1 K} \lra  S_{1}+K$ is more accurate.  
% This second network is obtained from the first by 
% adding the {\em intermediate} { $S_1 K$}.  (An ``intermediate'' was defined mathematically by Feliu and Wiuf~\cite{FeliuWiuf} to capture the concept of reaction-intermediates from chemistry.) It turns out, however, that adding this one intermediate has no qualitative effect on the dynamics, nor does making any reaction reversible.  
As an % other
example, we revisit the fully processive network in Eq.~\ref{eq:n-site-processive}.
% Informally, we can analyze network~\eqref{eq:n-site-processive} by
% first removing intermediates to obtain this network:
By iteratively removing intermediates $S_iK$ and $S_iF$  \cite{FeliuWiuf}, we
obtain the reduced network:
%  Based on \cite{FeliuWiuf} we can analyze network~\eqref{eq:n-site-processive} by
 % iteratively removing intermediates $S_iK$ and $S_iF$  until we
 % obtain the following network with known steady state properties:

\begin{equation} % Make first reversible.
  \label{eq:1-site}
  \begin{split}
  \xymatrix{
    S_0 + K  \ar @<.4ex> @{-^>} [r] ^-{}
    &\ar @{-^>} [l] ^-{} S_0 K \ar [r] ^-{}
    & S_n + K\\
    S_n + F \ar @<.4ex> @{-^>} [r] ^-{}
    &\ar @{-^>} [l] ^-{} S_{n} F \ar  [r] ^-{} 
    ^-{} 
    &S_0 + F
  }
\end{split}
\end{equation}
The network in Eq.~\ref{eq:1-site} is mathematically equivalent to the
single-site phosphorylation network (the $n=1$ case of the 
network in Eq.~\ref{eq:n-site-processive}), which is known to be globally
stable: each invariant set has a unique steady state, this steady
state is positive, and it is the global attractor of the invariant
set~\cite{AS}.  
%  As a consequence of the results described in \cite{FeliuWiuf} one
Thus, as explained above,
%from~ \cite{FeliuWiuf}, 
we conclude that the network in Eq.~\ref{eq:n-site-processive}
  has at least one (locally stable) steady state. % for some parameters,  in some invariant set. 
  
In fact, we can say more about the network in
Eq.~\ref{eq:n-site-processive}: because it reduces to the globally stable network in Eq.~\ref{eq:1-site}, it too is globally stable~\cite{ConradiShiu,EithunShiu}.  We proved this via results on how the dynamics are affected when intermediates are incorporated into a network. We turn to this topic next.

%  we can say more: using results on how the dynamics are affected when intermediates are incorporated into a network (the topic we turn to next), we proved, with Eithun, that the original network~\eqref{eq:n-site-processive} is globally stable~ \cite{ConradiShiu,EithunShiu}.
%: fully processive phosphorylation systems are globally stable \cite{ConradiShiu,EithunShiu}.

%by applying a result from
%  Sontag and Angeli \cite{AS} we were able to show, with Eithun,  that
 % the result from \cite{AS} holds for any $n\geq 1$: each invariant
 % set has a unique steady state, this steady state is positive, and it
 % is the global attractor of the invariant set \cite{ConradiShiu,EithunShiu}.

% It turns out that this is enough for us to draw the same conclusions about the original network~\eqref{eq:n-site-processive}: fully processive phosphorylation systems are globally stable \cite{ConradiShiu,EithunShiu}.

%\dblue{Are we ???}
%In this section, we describe how we draw the conclusions we mentioned in the above examples.  Specifically, we will highlight recent theorems that tell us how the dynamics and steady states are affected when intermediates are incorporated into a network -- and how such results yield further insights into multisite phosphorylation systems.

\subsection{Intermediates and dynamics}
As mentioned above, scientists have long used approximations of reaction systems % based on assumptions on fast intermediates
 (e.g., Michaelis-Menten), in which steady states correspond exactly to those of the original system --  but trajectories need not have such a correspondence.  %Recently, researchers have examined how well such reductions approximate the original system of interest.
In the case of a reaction system with intermediates, however, the trajectories are close to each other:
Cappelletti and Wiuf proved that the trajectories can be uniformly approximated on compact time intervals by the trajectories of the reduced reaction system obtained by removing intermediates~ \cite{approx-elim}.  
%Cappelletti and Wiuf, for instance, proved that the trajectory of can be uniformly approximated on compact time intervals by the trajectory of a reduced reaction system obtained by removing intermediates \cite{approx-elim}.  

Similarly, global stability of a network obtained by removing intermediates, or by making some reactions irreversible, can imply that the original system also is globally stable.  
%({\em Global stability} means that, for every choice of rate constants and conservation-law values, there is a unique steady state, and every trajectory converges to that steady state.)
Marcondes de Freitas {\em et al.} proved such a result, using monotone systems theory~\cite{Angeli2010}, for removing intermediates~\cite{Freitas1,Freitas2}; and Ali Al-Radwahi and Angeli proved such a result for making reactions irreversible~\cite{ali-angeli}.

\subsection{Future direction: intermediates and oscillations}
One focus of this section was on the question of how dynamics and steady states are affected when intermediates and/or reactions are added or removed.  What we presently can say about dynamics, however, is mostly limited to the situation of global stability.  We therefore are interested in other dynamics, such as oscillations.  

For instance, is the presence of a Hopf bifurcation preserved when reactions or intermediates are added or removed?  To pose a concrete question, we revisit the mixed-mechanism network in Eq.~\ref{eq:mixed-network}. Suwanmajo and Krishnan showed that it admits a Hopf bifurcation; {\em is this still true if any of the irreversible reactions is made reversible?}  Can the answer be explained by some general, yet-to-be-proven theorem that ``lifts'' Hopf bifurcations  when reactions or intermediates are added?

Some partial results in this direction were given recently by Banaji. %~\cite{banaji-inheritance}.  
Regarding the question posed above, Banaji proved that making any irreversible reactions reversible indeed preserves oscillations -- as do several other operations that add species or reactions~\cite{banaji-inheritance}.

%---------------------------------
%: PARAMETRIZATIONS
%---------------------------------
%\section{Steady-state parametrizations: understanding steady states in terms of catalytic constants} \label{sec:param}

%Ideas:
%\begin{enumerate}
%    \item TSS: \cite{MAPK, translated, signs, messi}
%    \item Rational parametrization theorem: \cite{TG}
%    \item \cite{a6maya}
%\end{enumerate}

%Possible questions to pose:
%\begin{enumerate}
 %   \item What conditions on the network (beyond being toric) guarantee that a parametrization exists?
  %  \item When the rational parametrization theorem holds, does it follow that there exists a parametrization in terms of the catalytic constants?
%\end{enumerate}

%---------------------------------
%: Dynamics
%---------------------------------
\section{Dynamics of multisite phosphorylation of a single substrate} \label{sec:dynamics}
This section highlights what is known -- and what is open -- about the dynamics of the dual-site phosphorylation systems in Figure \ref{fig:multisite-aims}.  
%(Here we will summarize what is known -- and what is open -- about the dynamics of multisite systems.  We will explain how the above topics and results contribute to our understanding of the dynamics.  And we will highlight open questions that our field is working toward resolving.)  
We are not attempting to be thorough -- there are many more multisite systems, and we point the  reader to~\cite{scaffold,Enzyme-sharing,compartments,jolley,kapuy,SK}.

  \subsection{Distributive $n$-site systems, including Figure \ref{fig:multisite-aims}(A)} \label{sec:a}
Most studies on the mathematics of multisite phosphorylation
have focused on %multisite phosphorylation under 
the case of a sequential and fully {\em distributive}
mechanism
 \cite{a6maya, Enzyme-sharing,KathaMulti,ManraiGuna,Markevich,MAPK,ortega,thomson2009unlimited,WangSontag,TSS}:
% Concretely, consider the network:
\begin{equation}
  \label{eq:distributive-network}
  \begin{split}
  \xymatrix{
    % DISTRIBUTIVE
    S_0 + K \ar @<.4ex> @{-^>} [r] ^-{}
    &\ar @{-^>} [l] ^-{} S_{0} K \ar %@<.4ex> @{-^>} 
    [r] ^-{{}} 
    &S_1 + K
    \ar @<.4ex> @{-^>} [r] ^-{}
    &\ar @{-^>} [l] ^-{}
    S_{1} K 
    \ar %@<.4ex> @{-^>} 
    [r]^-{} 
    & S_2 + K
    \ar @<.4ex> @{-^>} [r] ^-{}
    &\ar @{-^>} [l] ^-{}
    \cdots
    \ar %@<.4ex> @{-^>} 
    [r]^-{} 
    & S_n + K
    \\
    % DISTRIBUTIVE
    S_n+F 
    \ar @<.4ex> @{-^>} [r] ^-{}
    &\ar @{-^>} [l] ^-{}
    \cdots
    \ar @<.4ex> %@{-^>}
    [r]^-{} 
    & S_2 + F \ar @<.4ex> @{-^>} [r] ^-{}
    &\ar @{-^>} [l] ^-{} S_{2} F \ar %@<.4ex> @{-^>} 
    [r] ^-{{}} 
    &S_1 + F
    \ar @<.4ex> @{-^>} [r] ^-{}
    &\ar @{-^>} [l] ^-{}
    S_{1} F \ar %@<.4ex> @{-^>} 
    [r]^-{} 
    & S_0 + F
  }
  \end{split}
\end{equation}

This network is multistationary for $n \geq 2$; this result can be proven via a monomial
  parametrization of the steady states \cite{KathaMulti,TSS}.  
As for the maximum number of positive steady states, this number is at  most $2n-1$; this was proven by Wang and Sontag using an elimination similar to those in Section~\ref{sec:intermediates} \cite{WangSontag}.   A rationale to
  obtain rate constants for which the system has the
  maximum number of positive steady states was described in
  \cite{FHC}. 
%
%  For this system it has been shown that there are at most $2n-1$
%  positive steady states \cite{WangSontag} and a rationale to
%  obtain values for the rate constants where the system has the
 % maximal number of positive steady states has been described
%  \cite{FHC}. The biological significance of multiple {\em stable}
 % steady states has been described in, for example,
 % \cite{thomson2009unlimited}. The system admits a monomial
 % parameterization\cite{TSS} and multistationarity is straightforward
%  to establish\cite{KathaMulti,TSS}.
%
%\subsection{Distributive systems: Figure \ref{fig:multisite-aims}(A)} \label{sec:a}
%A similar steady-state analysis was performed by Wang and Sontag~\cite{WangSontag} for the fully distributive, dual-site phosphorylation network.  Their analysis of this network,
%which is referred to in Figure~\ref{fig:multisite-aims}(A) and which appears later in~\eqref{eq:distributive-network}, revealed that this network {\em is} multistationary: it admits up to 3 steady states (of which 2 may be stable~\cite{bistable}).
%
As for multiple {\em stable} steady states, the distributive {\em dual-site} network (the $n=2$ case of 
the network in Eq.~\ref{eq:distributive-network}) admits bistability~\cite{bistable}.  
%Also for this dual-site network, a Hopf bifurcation has been reported (not yet confirmed)~\cite{Errami}.  %Earlier results on multistationary (there are up to 3 steady states) by Wang and Sontag relied on elimination similar to those described in Section~\ref{sec:intermediates}~\cite{WangSontag}.
% 
%, and the set of steady states is parametrized by functions with monomial coordinates~\cite{translated,TG,TSS} -- {\color{blue} connect with earlier section!.}  

\subsection{Distributive systems, Michaelis-Menten approximation: Figure \ref{fig:multisite-aims}(B)} \label{sec:M-M} 
The Michaelis-Menten (MM) approximation of the distributive dual-site 
network %~\eqref{eq:distributive-network}
%dual-site network in Figure~\ref{fig:multisite-aims}(A) 
is a two-dimensional ODE system~\cite{WangSontag08}.  
The validity of this approximation for phosphorylation systems has been called into question~\cite{salazar}.  Nevertheless, analyzing this system is valuable,
 as dynamical properties of the MM approximation
sometimes can be ``lifted'' to the fully distributive system~\cite{WangSontag08}.  Indeed, the MM approximation is bistable, and Hell and Rendall ``lifted'' bistability to the full system~\cite{bistable}.

Oscillations, however, are precluded in the MM approximation, because every trajectory converges to some steady state.  This was proven by Wang and Sontag~\cite{WangSontag08} via monotone systems theory, and then re-proven by Bozeman and Morales \cite{no-osc} via the simpler Bendixson's Criterion.  

\subsection{Processive systems, including Figure \ref{fig:multisite-aims}(D)} \label{sec:pro-lit} 
As discussed in \S \ref{sec:intermediates},
 multisite phosphorylation systems that are fully {\em processive} 
 are globally stable~\cite{ConradiShiu,EithunShiu}, and thus, %~\cite{ConradiShiu}.   
%(This builds on Gunawardena~\cite{Guna} and Conradi~{\em et al.}~\cite{ConradiUsing}.)  
in contrast to distributive systems, do {\em not} admit bistability or oscillations.  
%
%For any mass-action kinetics system arising from the $n$-site, fully processive phosphorylation/dephosphorylation network, 
%each invariant set contains a unique steady state, which is a global attractor~\cite{ConradiShiu}.  %Additionally, we obtained a monomial parametrization of the steady states. %Our proofs relied on a technique of Johnston for using “translated” networks to study systems with “toric steady states”, sign conditions for injectivity of polynomial maps, and a result from monotone systems theory due to Angeli and Sontag.
%
This result revealed that processive systems lack the more interesting information-processing capabilities of distributive systems -- an idea that existed in the experimental literature, but, until recently, lacked a proof.

%Recently, Eithun and the second author extended the global convergence result to allow for more intermediates and for reactions that may be irreversible~\cite{EithunShiu}.  The proof uses results from monotone systems theory~\cite{Angeli2010} and network/graph reductions that correspond to removal of intermediates~\cite{Freitas2}.

%{\color{blue} Consider noting here that Aoki et al.'s processive network~\cite{Aoki} is really both processive and distributive.}

\subsection{The ERK-mechanism system: Figure \ref{fig:multisite-aims}(C)} \label{sec:c}
Figure \ref{fig:multisite-aims}(C) refers to the network of ERK regulation by dual-site phosphorylation by MEK and dephosphorylation by MKP3, here:% {\color{red} To do: Label $k_{\rm cat}$, $k_{\rm off}$,... below }
%\begin{displaymath}
\begin{equation}
  \label{eq:erk-1}
  \xymatrix@C=2ex{
    S_{00}+K 
    \ar @<.4ex> @{-^>} [r] ^-{}
    &
    \ar @{-^>} [l] ^-{}
    S_{00}K 
    \ar [r]^-{}
    & 
    S_{01}K
    \ar [r] ^-{k_{\rm cat}}
    &
    S_{11}+K 
    &  
    S_{11}+F   
    \ar @<.4ex> @{-^>} [r] ^-{}
    &
    \ar @{-^>} [l] ^-{}
    S_{11}F \ar [r] ^-{}
    &
    S_{10}F
    \ar [r] ^-{l_{\rm cat}}
    & 
    S_{00}+F
    \\
    S_{01} K
    \ar @<.4ex> @{-^>} [r] ^-{k_{\rm off}}
    &
    \ar @{-^>} [l] ^-{}
    S_{01}+K
    & 
    &
    &
    S_{10}F     
    \ar @<.4ex> @{-^>} [r] ^-{l_{\rm off}}
    &
    \ar @{-^>} [l] ^-{}
    S_{10}+F
    &
    &
    \\
    S_{10} + K     
    \ar @<.4ex> @{-^>} [r] ^-{}
    &
    \ar @{-^>} [l] ^-{}
    S_{10} K 
    \ar [r] ^-{}  
    &
    S_{11}+E   
    & 
    &
    S_{01} + F 
    \ar @<.4ex> @{-^>} [r] ^-{}
    &
    \ar @{-^>} [l] ^-{}
    S_{01}F  
    \ar [r] ^-{}
    &
    S_{00}+F~.
    &
  }
  \end{equation}
%\end{displaymath}
Here $S_{ij}$ represents the substrate with $i$ phosphate groups attached at the first site and $j$ at the second.  This notation reflects the fact that this network is {\em non-sequential}: the first site to be phosphorylated (via $ S_{00}K \to S_{01}K$) is also the first to be dephosphorylated ($S_{11}F \to S_{10}F$).

%Rubinstein {\em et al.}\ proved that 
The ERK network %~(\ref{eq:erk-1}--\ref{eq:erk-3}) 
admits bistability and oscillations~\cite{long-term}.  
However, when 
 $k_{\rm cat} >> k_{\rm off}$ and
 $l_{\rm cat} >> l_{\rm off}$, then the network reduces to 
a network {\em without} bistability or oscillations, namely, the fully processive network in Figure~\ref{fig:multisite-aims}(D). 
% the 
(That network comprises only the reactions in Eq.~\ref{eq:erk-1} above.)
 % processive dual-site phosphorylation 
%This limiting network is the one in Figure~\ref{fig:multisite-aims}(D): it does not admit bistability or oscillations. %:
% LIMIT IS THE PROCESSIVE NETWORK
%\begin{align}
%\label{eq:pro-dual-erk}
%S_{00}+E  \lra S_{00}E \to S_{01}E \to S_{11}+E~ \quad  \quad \quad
%S_{11}+F  \lra S_{10}F \to S_{10}F \to S_{00}+F~.
%\end{align}
%
%In fact, this processive network~\eqref{eq:erk-1} is globally stable: this is part of what Mitchell Eithun proved, mentioned above in \S \ref{sec:lift}, by generalizing the aforementioned global convergence result for processive systems \cite{ConradiShiu}.
%
%This result is obtained by modifying the argument in~\cite{ConradiShiu} for the phosphorylation/dephosphorylation version of network~\eqref{eq:pro}, and this differs mathematically only by a single reaction.
%
Accordingly, Rubinstein {\em et al.} posed the question, {\em How do bistability and oscillations in the ERK network emerge from the processive limit?}
More precisely, when the ``processivity levels''
$\frac{k_{\rm cat}}{k_{\rm cat} + k_{\rm off}}$ and $ \frac{l_{\rm cat}}{l_{\rm cat} + l_{\rm off}}$ are arbitrarily close to 1, is the ERK network still bistable and oscillatory?

%Here we rephrase this question, via {\em levels of processivity} $\pi_k:= \frac{k_{\rm cat}}{k_{\rm cat} + k_{\rm off}}$ and $\pi_p := \frac{l_{\rm cat}}{l_{\rm cat} + l_{\rm off}}$ that must lie between 0 and 1:

%\begin{problem} \label{q:level-proc}
%Is the ERK network~(\ref{eq:erk-1}--\ref{eq:erk-3}) with levels of processivity $\pi_k$ and $\pi_p$ %that are arbitrarily close to 1, bistable?   
%Is it oscillatory?
%\end{problem}

% MIXED 
\subsection{Mixed-mechanism systems: Figure \ref{fig:multisite-aims}(E)} \label{sec:e}
Only recently have there been studies of mixed mechanisms (partially
processive, partially distributive)~\cite{Aoki}. One such network, in
which phosphorylation is processive and dephosphorylation is
distributive, was depicted earlier in Eq.~\ref{eq:mixed-network}
(Example~\ref{ex:mixed}). % and~\ref{ex:mixed-continued}). 
% If we specify the total concentrations $\left( \Ktot,~\Ptot,~\Stot
% \right)$, one expects finitely many steady states. 
This network is {\em not} multistationary (Suwanmajo and Krishnan proved this 
via an elimination procedure like those described in
Section~\ref{sec:intermediates})~\cite{SK}. 
%Using an elimination procedure similar to what
%~\eqref{eq:param}, 
%Suwanmajo and Krishnan proved that network~\eqref{eq:mixed-network}
%does {\em not} admit more than one steady state, that is, it is {\em
%not} multistationary~\cite{SK}.
It follows, by a standard application of the Brouwer fixed-point
theorem, that there is always a unique steady state (for all rate
constants and conservation-law values).  This proves half of a
conjecture that we posed~\cite{ConradiShiu}.

We were surprised when the other half of our conjecture was disproven:
in contrast with processive systems (\S \ref{sec:pro-lit}), mixed
systems need {\em not} be globally stable.  In fact, the network in
Eq.~\ref{eq:mixed-network} exhibits oscillations~\cite{SK}!   
The significance of this result, as explained by its discoverers,
Suwanmajo and Krishnan, is that the network in
Eq.~\ref{eq:mixed-network} ``could be the simplest enzymatic
modification scheme that can intrinsically exhibit
oscillation''~\cite[\S 3.1]{SK}. Accordingly, this network's capacity
for oscillations is an interesting future direction. 
%Another candidate for the simplest such network is the ERK
%network~(\ref{eq:erk-1}--\ref{eq:erk-3}). 
%ERK-mechanism network in Figure \ref{fig:multisite-aims}(C).
%Accordingly, both networks' capacity for oscillations are an
%interesting future direction.

\begin{figure}[!h]
  \centering
  \includegraphics[width=0.9\linewidth]{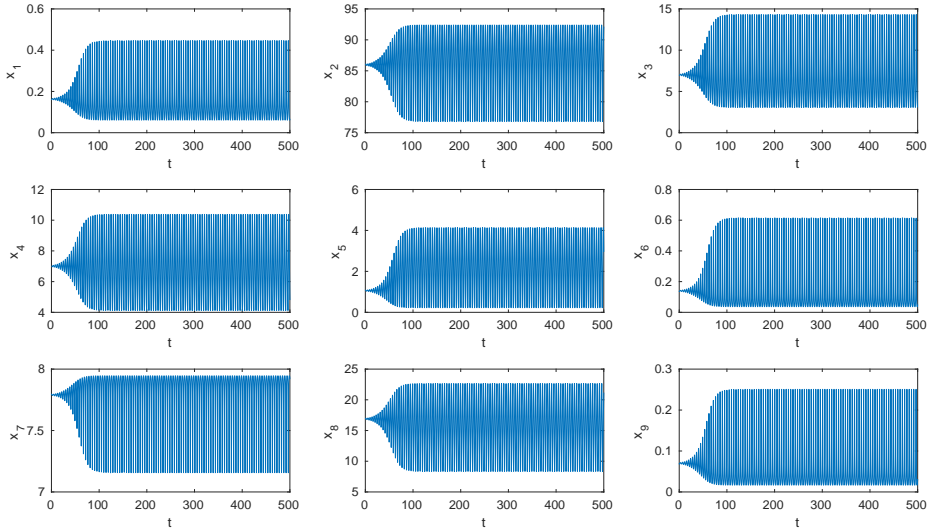}
  \caption{\label{fig:osci}
    Oscillations in the mixed-mechanism network (Eq.~\ref{eq:mixed-network}),
    where rate constants are from Table~\ref{tab:rcs}~\cite{SK},
    and, in contrast with \cite{SK}, 
%    While the parameter values
  %  are taken from \cite{SK}, different total concentrations and
    %initial values have been used: 
    $(\Ktot, \Ftot,\Stot)=(100,8,40)$ and the initial values are
    $x_1=0.16309$, $x_2=85.978$,
    $x_3=7.0112$, $x_4=7.0112$, $x_5=1.1$, $x_6=0.13971$,
    $x_7=7.7902$, $x_8=16.854$, $x_9= 0.070112$.
  } 
\end{figure}

  As a starting point,
  we display in Figure~\ref{fig:osci} such oscillations arising from the
  mixed-mechanism network.  The rate constants we used,
  from~\cite{SK}, are displayed in Table~\ref{tab:rcs}.
  Suwanmajo and Krishnan displayed oscillations for 
  $(\Ktot,\Ftot)=(17.5,5)$, while those in Figure~\ref{fig:osci} 
  used   $(\Ktot,\Ftot)=(100,8)$.  These results suggest that 
  oscillations exist over a wide range of $\Ktot$ -- and (possibly) $\Ftot$ -- values,
  extending what was found earlier when only $\Ktot$ was allowed to vary;~\cite{SK}.
%   we summarize the values of the rate constants
 % used by Suwanmajo and Krishnan in Table~\ref{tab:rcs} and use
 % Fig.~\ref{fig:osci} to demonstrate that periodic orbits exist over a
%  wide range of $\Ktot$- and (possibly) $\Ftot$-values: in \cite{SK},
%  oscillations are displayed for $\Ktot=17.5$ and $\Ftot=5$, while
%  those displayed in Fig.~\ref{fig:osci} have been obtained for
%  $\Ktot=100$ and $\Ftot=8$ (see caption of Fig.~\ref{fig:osci} for
%  details).

\begin{table}[!h]
  \centering
  \begin{tabular}{|c|c|c|c|c|c|c|c|c|c|} \hline
    $k_{1}$ &  $k_{2}$ &  $k_{3}$ & $k_{4}$ &  $k_{5}$ &  $k_{6}$ & $k_{7}$ & $k_{8}$ &  $k_{9}$ &  $k_{10}$ \\ \hline 
    1 & 1 & 1 & 1 & 100 & 1 & $0.9$ & 3 & 1 & 100 \\ \hline
  \end{tabular}
  \caption{\label{tab:rcs}
    Rate constants, from~\cite[Supplementary Information]{SK}, leading to oscillations 
   in the network in
Eq.~\ref{eq:mixed-network}. }
%    . Constants have
 %   been taken from the description of Fig.~4 in the supporting
  %  information of \cite{SK}.
\end{table}

%--------------------------------
%: End
%---------------------------------
\section{Future directions} \label{sec:future}
To add to the future directions already mentioned, here we highlight two more goals:
(1) characterizing bistability (so, going beyond multistationarity) and the corresponding parameter values, and (2) developing criteria for Hopf bifurcations that are well suited to PTM networks.  

Many researchers, ourselves included, have devoted much energy to
analyzing 
multistationarity in PTM networks
(and more generally in networks involving enzymes, substrates, and intermediates~\cite{messi})
and others~\cite{mss-review},
 but what we really are interested in
here, in terms of biology, is bistability.  Indeed, if an
experimentalist discovered a new PTM network and wanted to know what
dynamical properties it has, we have several techniques for assessing
whether this network is multistationary, but embarrassingly few for
determining whether it is bistable \cite{which-small}.  We would also
want to go further, to find witnesses for bistability (values of rate
constants, conservation-law values, and species concentrations at
which the system is bistable) and descriptions of the parameter-space
regions for which the system is bistable (or even just
multistationary).  Progress in this direction has been made for
    distributive dual-site and other PTM networks \cite{fein-024,fein-043}. Such
  developments have been 
 aided by techniques from areas of mathematics such as matroid theory~\cite{KathaMulti,messi}, degree theory~\cite{a6maya, CFMW} and numerical algebraic geometry~\cite{NAG-life-sci,case-study,decomposing}. We expect these and other areas to contribute in the future.

Turning our attention to oscillations, we need more tools for establishing that a network admits sustained oscillations or for precluding them. %well-suited to networks arising in biology
Returning to a question posed at the beginning of this Perspective, which PTM networks admit oscillations or at least a Hopf bifurcation?  This area has seen some recent progress~\cite{Errami,jolley,long-term,SK}, but deserves more attention.

%Ideas:
%\begin{enumerate}
 %   \item a conjectured upper bound on the number of steady states~\cite{WangSontag} was disproven in some cases~\cite{FHC}, yet the remaining cases are still open.  
 %   \item Steady-state invariants~\cite{Guna,ManraiGuna} -- for experiments -- one direction that the field is going?
 %   \item Maybe: \cite{scaffold}
%\end{enumerate}

%---------------------------------
%: End
%---------------------------------
\section{Conclusions} \label{sec:end}
What we hoped to convey in this Perspective is that the structure of a PTM network constrains its dynamics (Figure~\ref{fig:multisite-aims}).  
This is an old idea.  The question of how a (general) network constrains its dynamics was the main focus of the mathematical study of reaction systems (including chemical reaction network theory~\cite{FeinOsc}) in the years prior to the scope of this Perspective.  
In other words, the goal was to obtain results without knowing the rate constants.  

An updated version of this goal, 
given our interest in networks coming from biology, 
%, given the recent shift toward networks coming from biology (consider, for instance, the concept of ``absolute concentration robustness'' \cite{stoch-ACR,karp,sf10})), 
is to obtain results when not only are the rate constants unknown, but also aspects of the network structure, such as the precise enzymatic mechanism (Figure~\ref{fig:proc-dist}).  
Progress toward this goal has been aided by the methods highlighted in this Perspective --
including using steady-state parametrizations and results that clarify the effect of intermediates or extra reactions.  These techniques form the beginning of a theory of reaction systems well suited to any PTM or other biological signaling networks -- those we already know and also those yet to be discovered.

%Multisite phosphorylation systems are among the best-studied models in the reaction network community, but nonetheless many questions remain unresolved.  This is problematic, because disruptions to these systems have implications for human health, %~\cite{cohen-role,Anas}, (ALREADY CITED ABOVE)
%and to develop treatments in the future will require a quantitative understanding of these phosphorylation systems.  
%Moreover, phosphorylation is ubiquitous; to quote from~\cite{PM}: ``Approximately one-third of all eukaryotic proteins are modified by phosphorylation on serine, threonine, or tyrosine during their lifetime in the cell''.  (need to adapt or paraphrase the previous sentence)

%*******************************************************************
%Acknowledgements
%*******************************************************************
\subsection*{Acknowledgements} AS was supported by the NSF
  (DMS-1312473/DMS-1513364).
%  The authors thank ??Mercedes  for insightful comments.
The authors also thank three anonymous referees whose comments helped improve this work.

\bibliographystyle{plain}
% At the end, switch to this bib style...
% will require this line in BBL:
% \providecommand{\natexlab}[1]{#1}
% plus we must order them by numbers...
% (also, requires the style file biophysj.bst in the
% document directory)
%\bibliographystyle{biophysj}
\bibliography{phos.bib}

\end{document}